\documentclass[two column,preprintnumbers,amsmath,amssymb,bibnotes,prl]{revtex4}
\usepackage{graphicx}
\usepackage{dcolumn}
\usepackage{bm}
\usepackage{braket}
\usepackage{setspace}
\begin{document}
\title{Gigantic directional asymmetry of luminescence in multiferroic CuB$_2$O$_4$}
\author{S.\ Toyoda}
\affiliation{Department of Advanced Materials Science, University of Tokyo, Kashiwa  277-8561, Japan}
\author{N.\ Abe}
\affiliation{Department of Advanced Materials Science, University of Tokyo, Kashiwa  277-8561, Japan}
\author{T.\ Arima}
\affiliation{Department of Advanced Materials Science, University of Tokyo, Kashiwa  277-8561, Japan}
\hyphenpenalty=5000\relax
\exhyphenpenalty=5000\relax
\sloppy

\begin{abstract}
We report direction dependent luminescence (DDL), i.e., the asymmetry in the luminescence intensity between the opposite directions of the emission, in multiferroic CuB$_2$O$_4$. Although it is well known that the optical constants can change with the reversal of the propagation direction of light in multiferroic materials, the largest asymmetry in the luminescence intensity was 0.5 $\%$ so far. We have performed a measurement of photoluminescence with a He-Ne laser irradiation (633 nm). The luminescence intensity changes by about 70 $\%$ with the reversal of the magnetic field due to the interference between the electric dipole and magnetic dipole transitions. We also demonstrate the imaging of the canted antiferromagnetic domain structure of (Cu,Ni)B$_2$O$_4$ by using the large DDL.
\end{abstract}

\maketitle

Magnetoelectric (ME) effect, the induction of magnetization by electric field or the induction of electric polarization by a magnetic field, has been intensively investigated in multiferroic materials \cite{Review}. In addition to the static ME effect, multiferroic materials show novel optical phenomena, because the oscillating magnetic (electric) dipole moments can be induced also by electric (magnetic) fields of light. One typical example is non-reciprocal directional dichroism (NDD), a change in optical absorption with the reversal of the propagating direction of light (${\bm k}$).
Recent studies have revealed that a number of materials exhibit a large magnitude of NDD signal. \cite{Saito1,Saito2,Saito3,Saito4,Arima,YTakahashi,Kezs,Bord,Okamura,YTakahashi2,onewaythz,Kibayashi,BiFeO3}. In contrast, there have been few reports on direction dependent luminescence (DDL), the asymmetry in the luminescence intensity between the opposite directions of emission. The first experimental observation of the DDL was reported by Rikken $\it{et\ al.}$ in a chiral Eu(($\pm$)tfc)$_3$ complex \cite{Rikken}. The luminescence intensity of paramagnetic Eu$^{3+}$ ions depends on whether the direction of the emission is parallel or antiparallel to the external magnetic field direction (magneto-chiral dichorism, MChD). Shimada $\it{et\ al.}$ investigated the DDL in paramagnetic rare-earth ions in ferroelectric (Ba,Sr)TiO$_3$ and La$_2$Ti$_2$O$_7$ in a magnetic field ${\bm B}$, and found that the luminescence intensity was dependent on the sign of ${\bm k}\cdot ({\bm P}\times {\bm B})$ \cite{ShimadaBaTiO3,ShimadaLa2Ti2O7}. 

\begin{figure}
\includegraphics*[width=8.6cm]{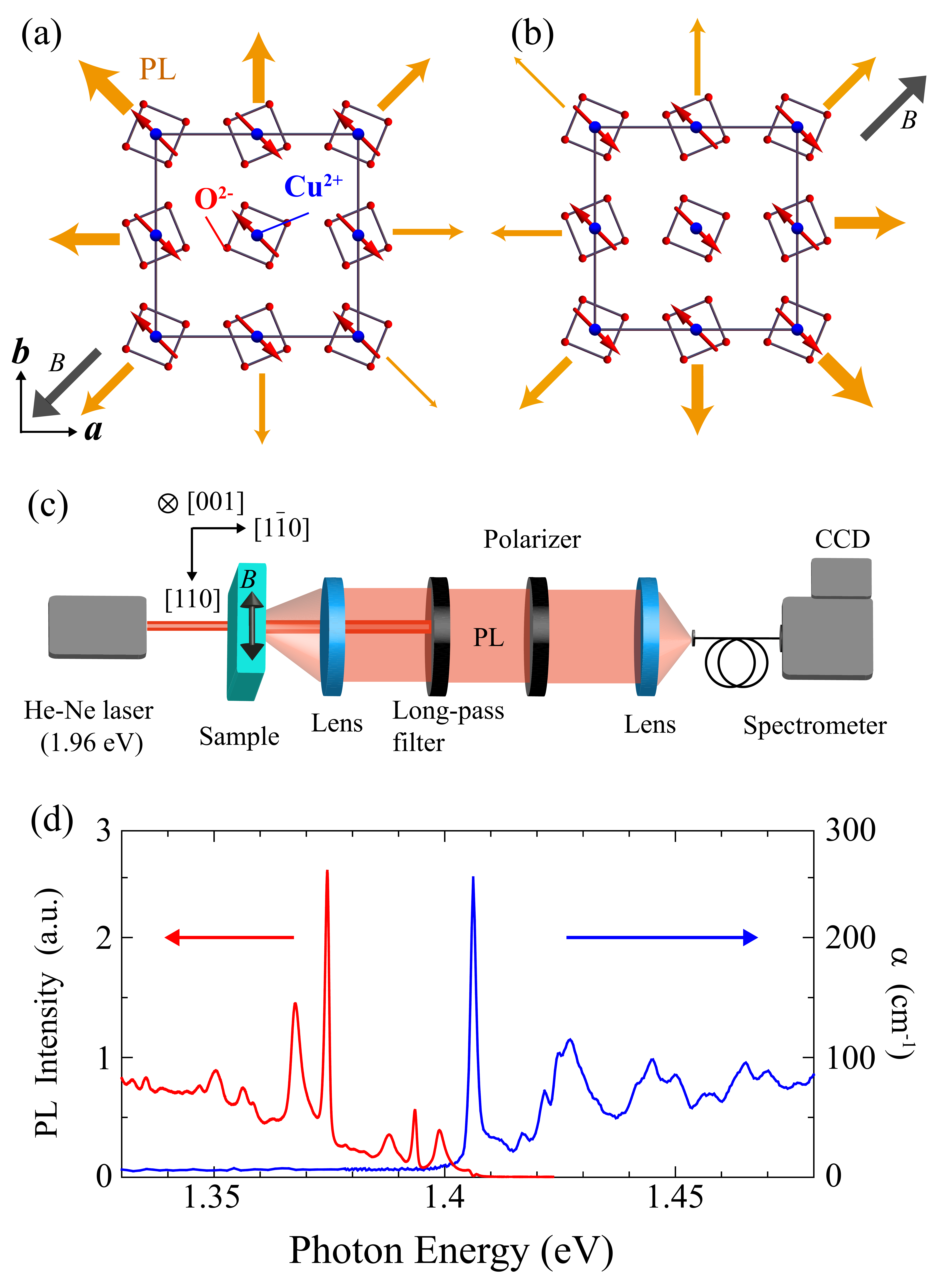}
\caption{
(color online). (a),(b) Schematic of direction dependent luminescence in CuB$_2$O$_4$ (orange arrows). Red arrows show the magnetic moments of Cu$^{2+}$ ions at $A$ sites in magnetic fields of (a) $B_{110}<0$ and (b) $B_{110}>0$. (c) Schematic illustration of the experimental setup. The magnetic field was applied along the [110] axis. The sample was excited by a 0.5 mW He-Ne laser at 633 nm (1.96 eV). The laser light was cut by a long-pass filter after the sample. The emitted light polarized along the [110] axis was selected by a polarizer and measured by a spectrometer. (d) Blue and red lines indicate optical spectra of absorption and PL for the light of $E^{\omega}\parallel[110]$ and $B^{\omega}\parallel [001]$ in zero magnetic field at $T=15$ K, respectively.
\label{fig:001}
}
\end{figure}
\begin{figure}
\includegraphics*[width=8.6cm]{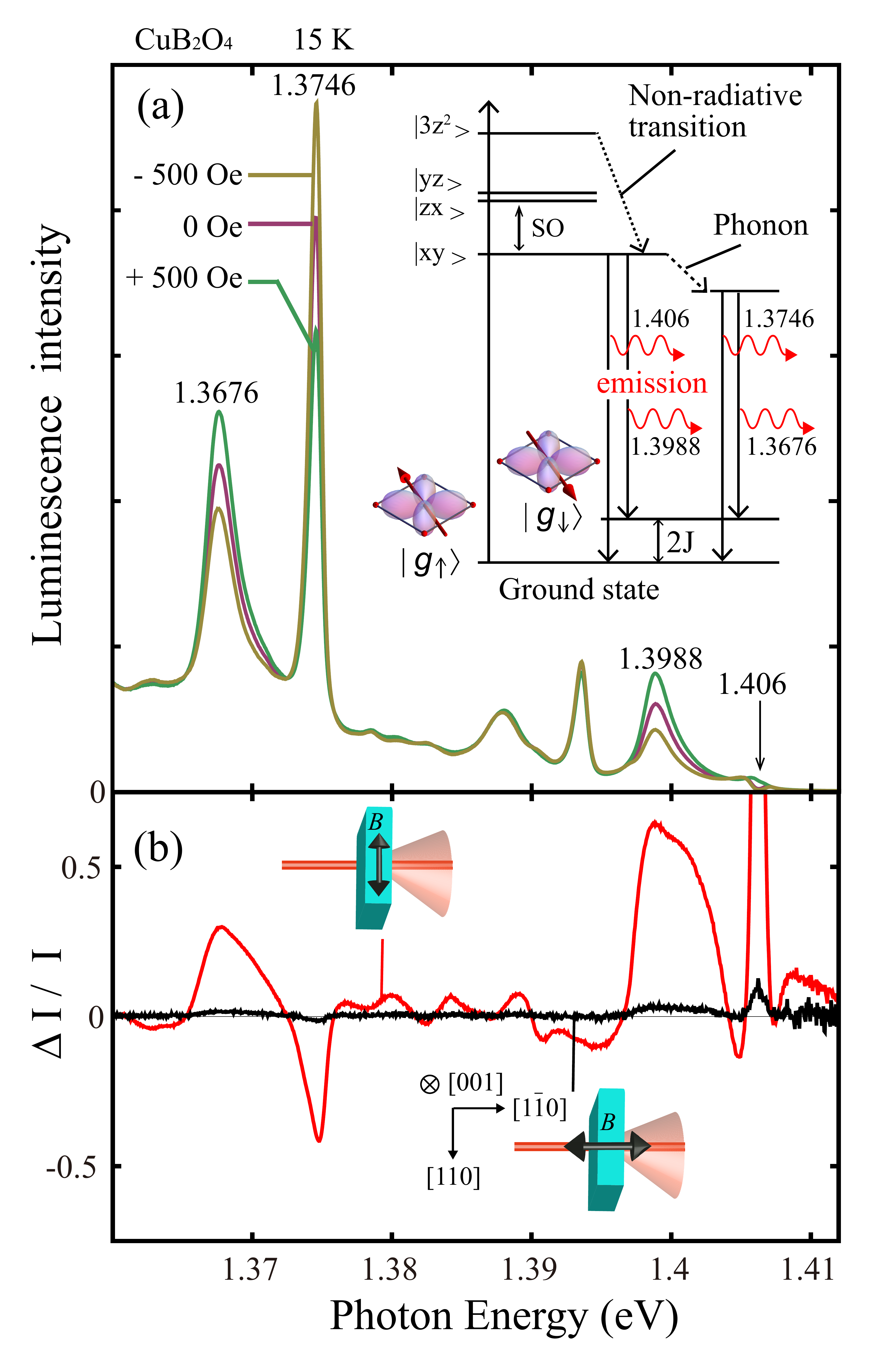}
\caption{
(color online). (a) Magnetic filed dependence of the PL spectrum at $T=15$ K in a magnetic field $B\parallel [110]$ for the emitted light of $E^{\omega} \parallel [110]$ and $B^{\omega} \parallel [001]$. The inset shows the energy diagram of Cu$^{2+}$ ions at A sites. The transition process is attached. (b) The DDL spectra ($\frac{\Delta I}{I}$) in the Voigt configuration (red) and the Faraday configuration (black).
\label{fig:002}
}
\end{figure}
\begin{figure*}
\includegraphics*[width=15cm]{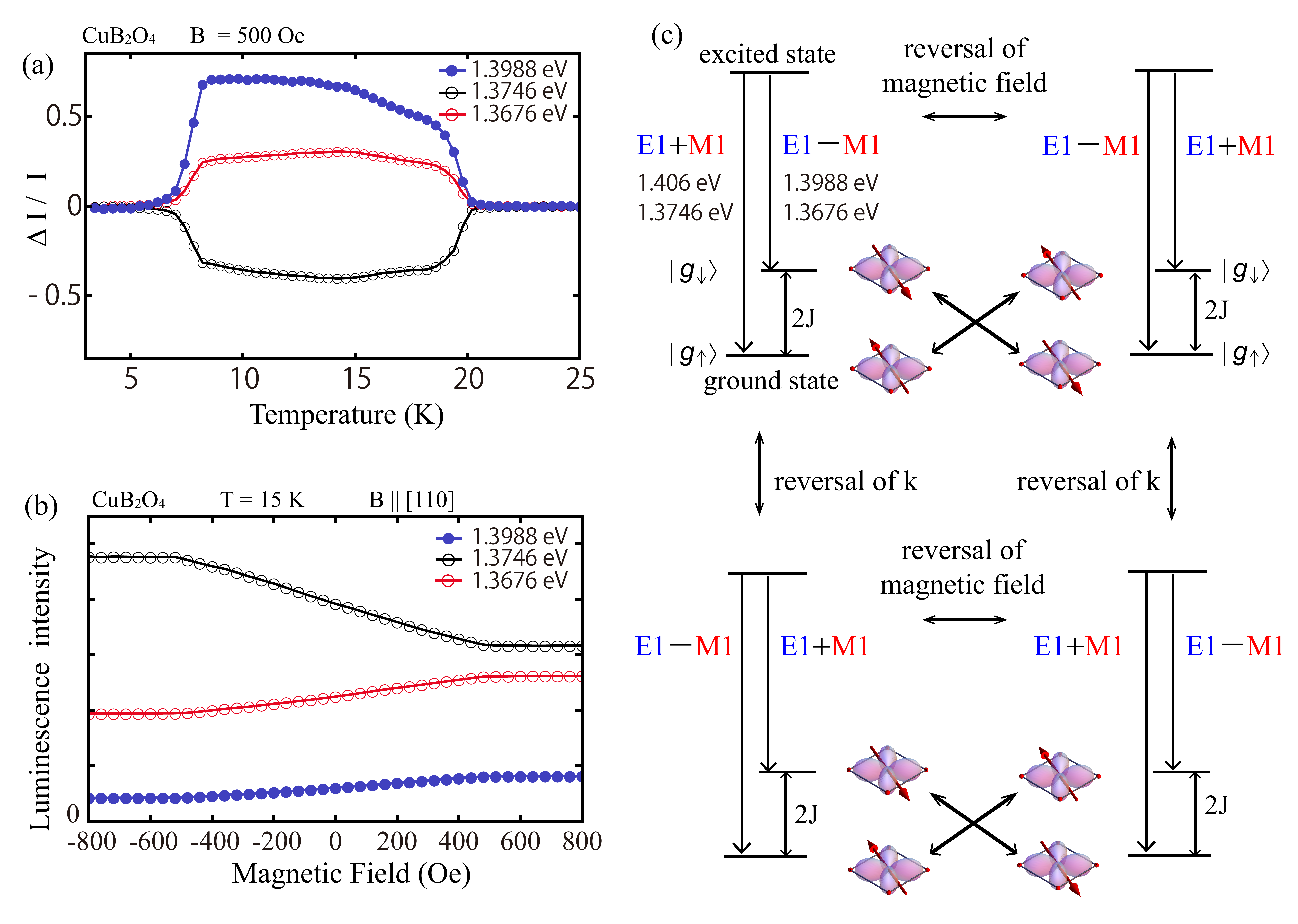}
\caption{
(color online). (a) Temperature dependence of the DDL signal $\frac{\Delta I}{I}$ at each peak position in a magnetic field of $B=500$ Oe.  (b) Magnetic-field dependence of the PL intensity at $T=15$ K in a magnetic field along the [110] axis. (c) Schematic illustration of the microscopic mechanism of the DDL. The sign of the interference term between the E1 and M1 transitions changes with the reversal of either the magnetic field or the propagating direction of light.
\label{fig:003}
}
\end{figure*}
To the best of our knowledge, the DDL has been reported only in the paramagnetic materials, and the magnitudes of the DDL asymmetry are smaller than 0.5 $\%$ \cite{ShimadaBaTiO3}. In this letter, we report DDL in a magnetically ordered noncentrosymmetric system. We performed a measurement of photoluminescence (PL) in multiferroic CuB$_2$O$_4$ with excitation by a He-Ne laser at 633 nm (1.96 eV). The observed DDL signal reaches 70 $\%$, which is about 100 times stronger than the previously reported value. Furthermore, we demonstrate that such a gigantic DDL gives a novel imaging technique for magnetic domain structures.

CuB$_{2}$O$_{4}$ crystallizes in a noncentrosymmetric tetragonal structure with a space group I$\bar{4}$2d \cite{crystal}. Cu$^{2+}$ ions occupy two inequivalent crystallographic sites denoted as $A$ and $B$. Cu$^{2+}$ ions at $A$ sites are surrounded by four O$^{2-}$ ions in planar square coordination with the site symmetry noncentrosymmetric $\bar{4}$. Cu$^{2+}$ ions at B site are surrounded by six O$^{2-}$ ions. The electronic configuration of Cu$^{2+}$ is $d^9$ (one hole) with $S=\frac{1}{2}$. The material undergoes successive magnetic transitions at $T_{N}=21$ K and $T^{\ast}=9$ K. Below $T^{\ast}$, magnetic moments at both $A$ and $B$ sites exhibit incommensurate helical order. Between $T_{N}$ and $T^{\ast}$, magnetic moments of Cu$^{2+}$ on $A$ sites exhibit commensurate canted antiferromagnetism, while magnetic moments of Cu$^{2+}$ $B$ site remain disordered \cite{Petra2,Boehm1,Boehm2}. In the canted antiferromagnetic phase, the weak ferromagentic moment can be rotated in the $[001]$ plane with the application of a weak magnetic field of $B=500$ Oe \cite{Petra2}. Figures\ \ref{fig:001}(a) and\ \ref{fig:001}(b) show the magnetic structures in external magnetic fields in the $[\bar{1}\bar{1}0]$ and $[110]$ directions, respectively. The material shows ME effect explained by the modification of the metal-ligand hybridization with Cu$^{2+}$ moments \cite{MetalLigand}. The electric polarization is induced along the $c$ axis in an external magnetic field along the [110] axis, and unchanged with the reversal of the magnetic field \cite{Khanh}. As a result, ${\bm P}\times {\bm M}$ appears in the $[1\bar{1}0]$ ($[\bar{1}10]$) direction in an external magnetic field in the $[\bar{1}\bar{1}0]$ ($[110]$) direction. Saito $\it{et\ al}$. reported gigantic NDD in this configuration for the near infrared light at 1.41 eV \cite{Saito1}, corresponding to the $d-d$ transition of a Cu$^{2+}$ hole at $A$ sites from the ground state $d_{x^{2}-y^{2}}$ to the excited state $d_{xy}$ \cite{SHG,Pisarev}. Here $x$, $y$, and $z$ denote the local coordinate axes at Cu $A$ sites, where $z$ is parallel to the crystallographic $c$ axis. The optical absorption coefficient changes by a factor of three with the reversal of a weak magnetic field of $B=500$ Oe. At a higher magnetic field, the material even shows one-way transparency of light, i.e., transparent for light propagating in one direction, while opaque for the light propagating in the opposite direction \cite{Oneway}. Such a material may also show the gigantic DDL effect.

Single crystals of CuB$_2$O$_4$ and (Cu$_{0.95}$Ni$_{0.05}$)B$_2$O$_4$ were grown by a flux method \cite{PetraSupple}. The crystals were oriented using Laue X-ray diffraction patterns. The thickness of each sample was 100 $\mu$m with the widest faces $(\bar{1}10)$. The sample was attached on a copper plate and cooled down with a closed-cycle refrigerator. We show in Fig.\ \ref{fig:001}(c) the experimental setup for the PL measurement. The sample was excited with a 0.5 mW He-Ne laser of wavelength 633 nm polarized along the $c$ axis. The emitted light polarized along the [110] axis was chosen by an analyzing prism, and the PL spectrum was measured by using a grating-type optical spectrometer and a CCD detector.

We show in Fig.\ \ref{fig:001}(d) the optical spectra of absorption and luminescence for the light of $E^{\omega}\parallel[110]$ and  $B^{\omega}\parallel [001]$ in zero magnetic field at $T=15$ K. The PL is observed below 1.406 eV, which corresponds to the zero-phonon absorption line at Cu $A$ sites, suggesting that the emitted light should originate from Cu $A$ sites. The zero phonon line at 1.406 eV is not strong for luminescence, because the emitted light is absorbed by the sample due to the large absorption peak. Figure\ \ref{fig:002}(a) shows magnetic field dependence of the PL spectrum for the emitted light of $E^{\omega}\parallel [110]$ and $B^{\omega}\parallel [001]$ in an external magnetic field $B\parallel [110]$ at $T=15$ K. The intensities of the luminescence peaks at 1.3988, 1.3746, and 1.3676 eV clearly change with the reversal of the magnetic field of $B=500$ Oe. We show in Fig.\ \ref{fig:002}(b) the spectrum of the DDL signal $\frac{\Delta I}{I}$ at $T=15$ K. Here, $I$ and $\Delta I$ denote the luminescence intensity in zero magnetic field and the change in the luminescence intensity with the reversal of the magnetic field of $B=500$ Oe, respectively. The DDL signal appears in the Voigt configuration (${\bm B}\perp {\bm k}\parallel {\bm P}\times {\bm M}$), while it disappears in the Faraday configuration (${\bm B}\parallel {\bm k}\perp {\bm P}\times {\bm M}$), as expected from the group theory. The observed $\frac{\Delta I}{I}$ signal at 1.406 eV is not attributed to the DDL but the gigantic NDD effect. The other three peaks at 1.3988, 1.3746, and 1.3676 eV are attributed to the DDL, because the material does not show any optical absorption in this region. The energy shift of these peaks from the zero-phonon line at 1.406 eV are 7.2, 31.4, and 38.4 meV, respectively. The luminescence at 1.406 and 1.3988 eV can be assigned to the transitions to $\ket{g_{\uparrow}}$ and $\ket{g_{\downarrow}}$, respectively. Here, $\ket{g_{\uparrow}}$ and $\ket{g_{\downarrow}}$ denote $\ket{x^2-y^2}$ with the opposite spin direction, as shown in the inset in Fig.\ \ref{fig:002}(a). The maximal energy gap between $\ket{g_{\uparrow}}$ and $\ket{g_{\downarrow}}$ is $2J$, where $J$ denotes the anitiferromagnetic exchange interaction. The observed energy shift of 7.2 meV well agrees with an inelastic neutron scattering study ($2J=7.7$ meV) \cite{Boehm3}. The spin-flopped state is however a linear combination of various one-magnon states. In fact, the luminescence peaks at 1.3988 and 1.3676 eV have a broad tail on the higher energy side. We assign the peaks at 1.3746 and 1.3676 eV to the transitions to $\ket{g_{\uparrow}}$ and $\ket{g_{\downarrow}}$ with one phonon, respectively, because the Raman scattering data suggests that there should be some lattice vibration modes around 30 meV \cite{PisarevRaman}. We also measured magnetic field dependence of the optical absorption. The propagating direction of light with the smaller absorption at 1.406 eV coincide with the emission direction with the larger luminescence intensity at 1.3988 and 1.3676 eV, i.e., the direction with the smaller luminescence intensity at 1.3746 eV.

We show in Fig.\ \ref{fig:003}(a) temperature dependence of the DDL signal $\frac{\Delta I}{I}$ at each peak position, measured in an external magnetic field $B$ of 500 Oe along the [110] axis. The gigantic DDL signal appears essentially in the canted antiferromagnetic phase. The signal disappears below $T^{\ast}=9$ K, corresponding to the transition from the canted antiferromagnetic phase to the helical phase. This evidently shows that the DDL should be ascribed to the magnetic order which breaks time reversal symmetry. Here one should note that the time reversal symmetry is revived in the low temperature helimagnetic phase. The DDL signal for the luminescence peak at 1.3988 eV is as large as $\frac{\Delta I}{I}=0.7$ at $T=10$ K, which is about 100 times stronger than the ever reported value in paramagnetic materials.

Next, we discuss the origin of the gigantic DDL signal. The DDL is understood in terms of the interference of the electric dipole (E1) and magnetic dipole (M1) transitions. According to Fermi's golden rule, the intensities of the emission $I_+$ and $I_-$ for opposite propagating directions are written as
\begin{equation}
I_{\pm}\propto\ |<g|\mathcal{H}_{E1} \pm \mathcal{H}_{M1}|e>|^2.
\end{equation} 
Here $e$ and $g$ represent the excited state and the ground state, respectively. $\mathcal{H}_{E1}$ and $\mathcal{H}_{M1}$ are the operators of the electric dipole and magnetic dipole transitions, respectively. Equation (1) explains that the interference of the E1 and M1 transitions results in the DDL. The previous research revealed that the excited state of Cu$^{2+}$ ions at $A$ sites $\ket{xy}$ should be hybridized with $\ket{yz}$ and $\ket{zx}$ via the spin-orbit coupling \cite{Oneway}. The transition from the ground state $\ket{x^2-y^2}$ to $\ket{xy}$ is M1 allowed, while that to $\ket{yz}$ and $\ket{zx}$ are E1 allowed. The E1 and M1 transition interfere with each other for the emission process, in the same way as the absorption process. 

We show in Fig.\ \ref{fig:003}(b) magnetic field dependence of the PL intensity at $T=15$ K. Although the intensities of the luminescence peaks at 1.3988 and 1.3676 eV increase with magnetic field, that at 1.3746 eV decreases. The result suggests that the E1 transition constructively (destructively) interfere with the M1 transition for the transitions to $\ket{g_{\uparrow}}$ ($\ket{g_{\downarrow}}$) for the emission propagating in $[1\bar{1}0]$ direction, while they destructively (constructively) interfere for the emission propagating in the opposite direction. The luminescence intensities for the transition to $\ket{g_{\uparrow}}$ for the two propagation vectors are written as
\begin{equation}
I_{\pm}^{g_{\uparrow}}\propto\ \sum_{\sigma=\uparrow,\downarrow} |<g_{\uparrow}|\mathcal{H}_{E1} \pm \mathcal{H}_{M1}|e_{\sigma}>|^2,
\end{equation} 
while those to $\ket{g_{\downarrow}}$ are given by
\begin{eqnarray}
I_{\pm}^{g_{\downarrow}}&\propto&\ \sum_{\sigma=\uparrow,\downarrow} |<g_{\downarrow}|\mathcal{H}_{E1} \pm \mathcal{H}_{M1}|e_{\sigma}>|^2 \nonumber \\
&=&\ \sum_{\sigma=\uparrow,\downarrow} |<{\it \Theta} g_{\uparrow}|\mathcal{H}_{E1} \pm \mathcal{H}_{M1}|{\it \Theta} e_{\sigma}>|^2\\
&=&\ \sum_{\sigma=\uparrow,\downarrow} |<g_{\uparrow}|{\it \Theta}^{-1} (\mathcal{H}_{E1} \pm \mathcal{H}_{M1}){\it \Theta} |e_{\sigma}>^{\ast}|^2 \nonumber \\
&=&\ \sum_{\sigma=\uparrow,\downarrow} |<g_{\uparrow}|(\mathcal{H}_{E1} \mp \mathcal{H}_{M1})|e_{\sigma}>|^2, \nonumber
\end{eqnarray}
where $e_{\uparrow}$ and $e_{\downarrow}$ denote the excited states with the magnetic moment parallel and antiparallel to the magnetic field direction, respectively. ${\it \Theta}$ is time-reversal operator. Equations (2) and (3) explain the experimental result, where the M1 and E1 transitions constructively (destructively) interfere for the transition to $\ket{g_{\uparrow}}$, when they destructively (constructively) interfere for that to $\ket{g_{\downarrow}}$.
By the reversal of ${\bm k}$, the relative relation between the E1 and M1 transitions is flipped, as shown in Fig.\ \ref{fig:003}(c). 

\begin{figure}
\includegraphics*[width=8.6cm]{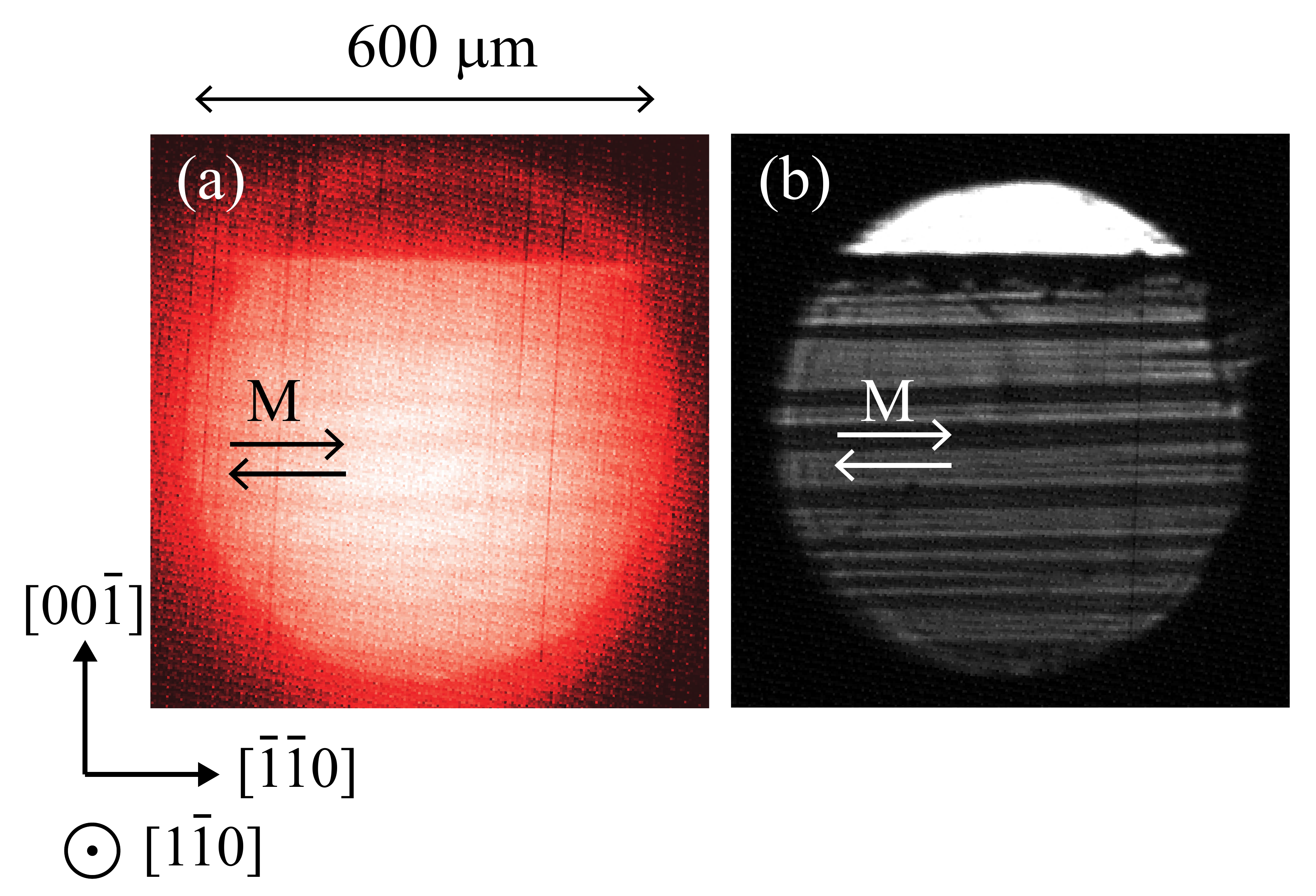}
\caption{
(color online). (a),(b) Magnetic domain images of a (Cu$_{0.95}$Ni$_{0.05})$B$_2$O$_4$ crystal obtained at $T=10$ K by (a) PL at 1.3988 eV and (b) absorption at 1.406 eV. The exposure times of both images are 200 ms. Bright and dark regions correspond to the magnetic domains with the magnetization in the $[110]$ and $[\bar{1}\bar{1}0]$ directions, respectively. The diameter of the hole of the copper plate is 600 $\mu$m. There is no sample in the top dark (bright) region in (a) ((b)).
\label{fig:004}
}
\end{figure}

We have performed measurements of both PL image and optical transmission image in (Cu$_{0.95}$Ni$_{0.05})$B$_2$O$_4$. We have chosen this material, because the magnetic domain size is much larger than CuB$_2$O$_4$, which is an advantage for the optical imaging. Figure\ \ref{fig:004}(a) displays the PL image taken at $T=10$ K after zero field cooling. The PL image was gained by using the luminescence peak around 1.3988 eV. Other luminescence lights were cut by short-pass filter. The magnetic domain structure is resolved as the contrast of the luminescence intensity. Figure\ \ref{fig:004}(b) shows the transmission image obtained by the absorption at 1.406 eV, where the gigantic NDD is observed. We have confirmed that the PL image agrees with the transmission image. We note that there is no domain structure observed above $T_N$ nor below $T^{\ast}$, as expected. This result confirms that the observed structure arises from the magnetic domain.

In conclusion, we have observed the gigantic DDL in CuB$_2$O$_4$. We succeeded in explaining the large DDL signal in terms of the interference between the electric dipole and magnetic dipole transitions. We have demonstrated that such nonreciprocal emission can be applied to the visualization of the magnetic domain structure. Since the NDD is reported in a number of materials, the DDL is not limited to CuB$_2$O$_4$, but may be also found in other multiferroic materials. The technique gives a novel tool to study magnetic domain structures of multiferroic materials which have large impacts in the field of spintronics.

This work was supported by Grant-in-Aid for JSPS Fellows (14J06840) and a Grant-in-Aid for Scientific Research from JSPS, JAPAN (24244045). S.T. acknowledges the financial support by JSPS through Program for Leading Graduate Schools (MERIT) and a research fellowship for young scientists. 

\end{document}